\documentclass[12pt,a4paper,onecolumn,oneside]{IEEEtran}
\IEEEoverridecommandlockouts
\usepackage{amsfonts,booktabs}
\usepackage{amsmath}
\usepackage{graphicx}
\usepackage{subfigure}
\usepackage{times}
\usepackage{geometry}
\let\oldbibliography\thebibliography
\renewcommand{\thebibliography}[1]{%
  \oldbibliography{#1}%
  \setlength{\itemsep}{0pt}%
}
\renewcommand{\baselinestretch}{1.65}

\newenvironment{remark}         {\begin{sideremark}\rm}{\end{sideremark}}
\newtheorem{sideremark}{Remark}

\newtheorem{lemma}{Lemma}
\newtheorem{theorem}{Theorem}

\def\reals{{\rm I\kern-.17em R}}
\def\nats{{\rm I\kern-.17em N}}

\newcommand{\beq}{\begin{equation}}
\newcommand{\eeq}{\end{equation}}
\newcommand{\beqa}{\begin{eqnarray}}
\newcommand{\eeqa}{\end{eqnarray}}

\begin{document}

\title{Power Allocation for Outage Minimization in Cognitive Radio Networks with Limited
Feedback}

\author{YuanYuan He and Subhrakanti Dey, \IEEEmembership{Senior Member,~IEEE}
\\ Department of Electrical and Electronic Engineering\\ University of Melbourne,
Vic. 3010, Australia\\ e-mail: \{yyhe, sdey\}@ee.unimelb.edu.au}
\maketitle
\thispagestyle{empty}
\vspace{-2cm}
\begin{abstract}
We address an optimal transmit power allocation problem that
minimizes the outage probability of a secondary user (SU) who is
allowed to coexist with a primary user (PU) in a narrowband
spectrum sharing cognitive radio network, under a long term
average transmit power constraint at the secondary transmitter
(SU-TX) and an average interference power constraint at the
primary receiver (PU-RX), with quantized channel state information
(CSI) (including both the channels from SU-TX to SU-RX, denoted as
$g_1$ and the channel from SU-TX to PU-RX, denoted as $g_0$) at
the SU-TX. The optimal quantization regions in the vector channel
space is shown to have a 'stepwise' structure. With this
structure, the above outage minimization problem can be explicitly
formulated and solved by employing the Karush-Kuhn-Tucker (KKT)
necessary optimality conditions to obtain a locally optimal
quantized power codebook. A low-complexity near-optimal quantized
power allocation algorithm is derived for the case of large number
of feedback bits. More interestingly, we show that as the number
of partition regions approaches  infinity, the length of interval
between any two adjacent quantization thresholds on the $g_0$ axis
is asymptotically equal when the average interference power
constraint is active. Similarly, we show that when the average
interference power constraint is inactive, the ratio between any
two adjacent quantization thresholds on the $g_1$ axis becomes
asymptotically identical. Using these results, an explicit expression for the
asymptotic SU outage probability at high rate quantization (as the
number of feedback bits goes to infinity) is also provided, and is
shown to approximate the optimal outage behavior extremely well
for large number of bits of feedback via numerical simulations.
Numerical results also illustrate that with 6 bits of feedback,
the derived algorithms provide SU outage performance very close to
that with full CSI at the SU-TX.
\end{abstract}
 \IEEEpeerreviewmaketitle
 \vspace*{-1cm}
 \section{Introduction}
\indent Scarcity of available vacant spectrum is limiting the
growth of wireless products and services \cite{Peha2009}.
Traditional spectrum licensing policy forbids unlicensed users to transmit in order to avoid unfavorable
interference at the cost of spectral utilization efficiency. This led to the idea of cognitive radio (CR) technology, originally introduced by J. Mitola \cite{Mitola99}, which holds tremendous promise to dramatically improve the efficiency of spectral utilization.\\
\indent The key idea behind CR is that an unlicensed/secondary
user (SU) is allowed to communicate over a frequency band
originally licensed to a primary user (PU), as long as the
transmission of SU does not generate unfavorable impact on the
operation of PU in that band. Effectively, three categories of CR
network paradigms have been proposed: interweave, overlay, and
underlay \cite{Goldsmith09}. In the underlay systems, also known
as spectrum sharing model, which is the focus of this paper, the
SU can transmit even when the PU is present, but the transmitted
power of SU should be controlled properly so as to ensure that the
resulting interference does not degrade the received signal
quality of PU to an undesirable level \cite{Kang09} by imposing
the so called interference temperature \cite{Ghasemi07}
constraints at PU
(average or peak interference power (AIP/PIP) constraint) and as well as to enhance the performance of SU transmitter (SU-TX) to SU receiver (SU-RX) link. \\
\indent Various notions of capacity for wireless channels include
ergodic capacity (for delay-insensitive services), delay-limited
capacity and outage probability (for real-time applications).
These information theoretic capacity notions constitute  important
performance measures in analyzing the performance limits of CR
systems. In \cite{Ghasemi07}, the authors investigated the ergodic
capacity of such a dynamic narrowband spectrum sharing model under
either AIP or PIP constraint at PU receiver (PU-RX) in various
fading environments. The authors of \cite{Suraweera08} extended
the work in \cite{Ghasemi07} to asymmetric fading environments. In
\cite{Musavian09}, the authors studied optimum power allocation
for three different capacity notions under both AIP and PIP
constraints. In \cite{Kang09}, the authors also considered the
transmit power constraint at the SU-TX and investigated the
optimal power allocation strategies to achieve the ergodic
capacity and outage capacity of SU under various combinations of
secondary transmit (peak/average) power constraints and
interference (peak/average) constraints.\\
\indent Achieving the optimal system performance requires the
SU-TX to acquire full channel state information (CSI) including
the channel information from SU-TX to PU-RX and the channel
information from SU-TX to SU-RX. Most of the above results assume
perfect knowledge or full CSI, which is very difficult to
implement in practice, especially the channel information from
SU-TX to PU-RX without PU's cooperation. A few recent papers have
emerged that address this concern by investigating performance
analysis with various forms of partial CSI at SU-TX, such as noisy
CSI and quantized CSI. With assumption of  perfect knowledge of
the CSI from SU-TX to SU-RX channel, \cite{Musavian092} studied
the effect of imperfect channel information of the SU-TX to PU-RX
channel under AIP or PIP constraint by considering the channel
information from SU-TX to the PU-RX as a noisy estimate of the
true CSI and employing the so-called 'tifr' transmission policy.
Another recent work \cite{suraweera09} also considered imperfect
CSI of the SU-TX to PU-RX channel in the form of noisy channel
estimate (a range from near-perfect to seriously flawed estimates)
and studied the effect of using a midrise uniformly quantized CSI
of the SU-TX to PU-RX channel, while also assumed the SU-TX had
full knowledge of the CSI from SU-TX to SU-RX channel. Recently,
\cite{KRZhang09} has proposed a practical design paradigm for
cognitive beamforming based on finite-rate cooperative feedback
from the PU-RX to the SU-TX and cooperative feedforward from the
SU-TX to the PU-RX. A robust cognitive beamforming scheme was also
analyzed in \cite{lan_zhang_09}, where full channel information on
SU-TX to SU-RX channel was assumed, and the imperfect channel
information on the SU-TX to PU-RX channel was modelled using an
uncertainty set. Finally, \cite{marques_giannakis09} studied the
issue of channel quantization for resource allocation via the
framework of utility maximization in OFDMA based CR networks, but
did not investigate the joint channel partitioning and rate/power
codebook design problem. The absence of a rigorous and systematic
design methodology for quantized resource allocation algorithms in
the context of cognitive radio networks motivated our earlier work
\cite{he_dey_tcomm_11}, where we addressed an SU ergodic capacity
maximization problem in a wideband spectrum sharing scenario with
quantized information about the vector channel space involving the
SU-TX to SU-RX channel and the SU-TX to PU-RX channel over all
bands, under an average transmit power constraint at the SU-TX and
an average interference constraint at the PU-RX. A slightly
different approach was taken in \cite{huang_plink_11,
eswaran_plink_11} where the SU overheard the PU feedback link
information and used this to obtain information about whether or
not the PU is in outage and
how the SU-TX should control its power to minimize interference on the PU-RX. \\
\indent In this paper, we  address the problem of minimizing the
SU outage probability under an average transmit power (ATP)
constraint at the SU-TX and an average interference power (AIP)
constraint at the PU-RX. Similar to \cite{he_dey_tcomm_11}, we
consider an infrastructure-based narrowband spectrum sharing
scenario where a SU communicates to its base station (SU-BS) on a
narrowband channel shared with a PU communicating to its receiver
PU-RX contained within the primary base station (PU-BS). The key
problem is the jointly designing the optimal partition regions of
the vector channel space (consisting of the SU-TX to SU-RX channel
(denoted  by power gain $g_1$) and the interfering channel between
the SU-TX and PU-RX (denoted by power gain $g_0$)) and the
corresponding optimal power codebook, and is solved offline at a
central controller called the CR network manager as in
\cite{he_dey_tcomm_11}, based on the channel statistics. The CR
network manager is assumed to be able to obtain the full CSI
information of the vector channel space $(g_1, g_0)$ in real-time
from the SU-BS and PU-BS, respectively, possibly via wired links
(similar to backhaul links in multicell MIMO networks connecting
multiple base stations). This real-time channel realization is
then assigned to the optimal channel partition and the
corresponding partition index is sent to the SU-TX (and to the
SU-RX for decoding purposes) via a finite-rate feedback link. The
SU-TX then uses the power codebook element associated with this
index for data transmission. It was shown in
\cite{he_dey_tcomm_11} that without the presence of the CR network
manager, and thus without the ability to jointly quantize the
combined channel space, the SU capacity performance is
significantly degraded if one carries out separate quantization of
$g_1$ and $g_0$. Even if such a CR network manager cannot be
implemented in practical cognitive radio networks due to resource
constraints, the results derived in this paper will serve as a
valuable benchmark. Under these networking assumptions, we prove a
'stepwise' structure of the optimal channel partition regions,
which  helps us explicitly formulate the outage minimization
problem and  solve it using the corresponding Karush-Kuhn-Tucker
(KKT) necessary optimality conditions. As the number of feedback
bits go to infinity, we show that the power level for the last
region approaches  zero, allowing us to derive a useful
low-complexity suboptimal quantized power allocation algorithm
called 'ZPiORA' for high rate quantization. We also derive some
other useful properties related to the channel quantizer structure
as the number of feedback bits approaches infinity: (a) under an
active AIP constraint, the length of interval between any two
adjacent quantization thresholds on $g_0$ axis is asymptotically
the same, and (b) while when the AIP is inactive, the ratio
between any two adjacent quantization thresholds on $g_1$ axis
asymptotically becomes identical. Finally, with these properties,
we derive explicit expressions for asymptotic (as the number of
feedback bits increase) behavior of the SU outage probability with
quantized power allocation for large resolution quantization.
Numerical studies illustrate
that with only 6 bits of feedback, the designed optimal algorithms provide secondary outage probability very close to that achieved by full CSI. With 2-4 bits of feedback, ZPiORA provides a comparable performance, thus making it an attractive choice for large number of feedback bits case. Numerical studies also show that ZPiORA performs better than two other suboptimal algorithms constructed using existing approximations in the literature. Finally, it is also shown that the derived asymptotic outage behavior approximates  the optimal outage extremely well as the  number of feedback bits becomes large.\\
\indent This paper is organized as follows. Section \ref{us1}
introduces the system model and the problem formulation based on
the full CSI assumption. Section \ref{us2} presents the joint
design of the optimal channel partition regions and an optimal power codebook
algorithm. A low-complexity suboptimal quantized power allocation strategy
 is also derived using novel interesting properties of the quantizer structure and optimal quantized power
codebooks.   In
Section \ref{us3}, the asymptotic behavior of SU outage
probability for high resolution quantization is investigated.
Simulation results are given in Section \ref{us4}, followed by
concluding remarks in Section \ref{us5}.
\section{System Model and Problem Formulation}\label{us1}
\indent We consider an infrastructure-based spectrum sharing
network where a SU communication uplink to the SU-BS coexists with
a PU link (to the PU-BS) within  a narrowband channel. Regardless
of the on/off status of PU, the SU is allowed to access the band
which is originally allocated to PU, so long as the impact of the
transmission of SU does not reduce the received signal quality of
PU below a prescribed level. All channels here are assumed to be
Rayleigh block fading channels. Let $g_1=|h_1|^2$ and
$g_{0}=|h_{0}|^2$, denote the nonnegative real-valued
instantaneous channel power gains for the links from SU-TX to
SU-RX and SU-TX to PU-RX respectively (where $h_1$ and $h_{0}$ are
corresponding complex zero-mean circularly symmetric channel
amplitude gains).
 The exponentially distributed channel power gain $g_1$ and
$g_{0}$, are statistically mutually independent and, without loss
of generality ({\em w.l.o.g}), are assumed to have unity mean. The
additive noises for each channel are independent Gaussian random
variables with, {\em w.l.o.g}, zero mean and unit variance. For
analytical simplicity, the interference from the primary
transmitter (PU-TX) to SU-RX is
neglected following previous work such as \cite{Kang09,Ghasemi07}(in the
case where the interference caused by the PU-TX at the SU-RX is
significant, the SU outage probability results derived in this paper
can be taken as lower bounds on the actual outage under
primary-induced interference). This assumption is justified when either the SU is outside PU's transmission range or the SU-RX is
equipped with interference cancellation capability particularly when the PU signal is strong.
\\
%\begin{figure}[h]
%\centering
%\includegraphics[scale=0.3]{g1}
%\caption{System model for a narrow-band spectrum-sharing
%scenario.} \label{g1}
%\end{figure}
\indent Given a channel realization (${g}_0, g_1$), let the
instantaneous transmit power (with full CSI) at the SU-TX be represented by $p({g}_0,
g_1)$, then the maximum mutual information of the SU for this
narrowband spectrum sharing system can be expressed as $R(g_1,p({g}_0, g_1))=\frac{1}{2}\log(1+g_1p({g}_0, g_1))$, where $\log$ represents the natural logarithm. The outage
probability of SU-TX with a pre-specified transmission rate $r_0$,
is given as, $P_{out}=Pr\{R(g_1,p({g}_0, g_1))<r_0\}$, where $Pr\{A\}$ indicates the probability of event $A$ occurring. Using the interference
temperature concept in \cite{Ghasemi07}, a common way to protect
PU's received signal quality is by imposing either an average or a
peak interference power (AIP/PIP) constraint at the PU-RX. %We had
%studied the optimization problem of maximizing SU capacity under
%both average transmit power constraint (ATP) at SU and AIP
%constraints at PUs with perfect CSI case and quantized CSI case
%respectively in \cite {paper2010}.
In \cite{Zhang09},  it was demonstrated  that the AIP constraint
is more flexible and favorable than the PIP constraint in the
context of transmission over fading channels. Let $Q_{av}$ denotes the average interference power limit
tolerated by PU-RX, then the AIP constraint can be written as,
$E[g_{0}p( {g}_0, {g}_1)]\leq Q_{av}$.\\
 \indent The following  optimal
power allocation problem that minimizes the outage probability of
SU in a narrowband  spectrum sharing with one PU, under both a
long term average transmit power (ATP) constraint at SU-TX and an
AIP constraint at the PU-RX, was considered in \cite{Kang09}
\begin{eqnarray}
&&\min_{p( {g}_0, g_1) \geq 0} ~~Pr\{\frac{1}{2}\log(1+g_1p({g}_0, g_1))<r_0\}\nonumber\\
&&~~~~s. t. ~~~~~E[p( {g}_0, {g}_1)]\leq P_{av},~~~~~E[g_{0}p( {g}_0, {g}_1)]\leq Q_{av}
\label{Q1}
\end{eqnarray}
where $P_{av}$ is the maximum average transmit power at SU-TX.\\
\indent With the assumption that perfect CSI of both $g_0$ and
$g_1$ is available at the SU-TX, the optimal power allocation
scheme for Problem (\ref{Q1}) is given by \cite{Kang09}:
$p^*({g}_0, {g}_1)= \frac{c}{g_1}$ when $\lambda^*_f+\mu^*_fg_0<\frac{g_1}{c}$, and $0$ otherwise,
where $c=e^{2r_0}-1$, and $\lambda^*_f$, $\mu^*_f$ are the optimal
nonnegative Lagrange multipliers associated with the ATP constraint and the AIP constraint, respectively, which can be obtained by solving $\lambda^*_f(E[p({g}_0, {g}_1)]-P_{av})=0$ and  $\mu^*_f(E[g_{0}p( {g}_0, {g}_1)]-Q_{av})=0$ .\\
\indent However, the assumption of full CSI at the SU-TX
(especially that of ${g}_0$) is usually unrealistic and difficult to implement in
practical systems, especially when this channel is not time-division duplex (TDD). In the next section, we are therefore
interested in designing a power allocation strategy of the outage probability minimization Problem (\ref{Q1}) based on quantized CSI
at the SU-TX acquired via a no-delay and error-free feedback link with limited
rate.
\section{Optimum Quantized power allocation (QPA) with imperfect $g_1$ and ${g}_0$ at SU-TX}\label{us2}
\subsection{Optimal QPA with limited rate feedback strategy}
\begin{figure}[h]
\centering
\includegraphics[scale=0.8]{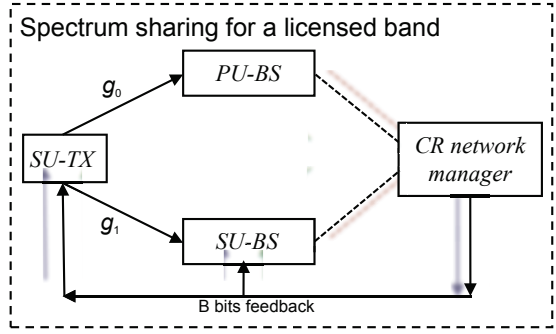}
\caption{System model for narrowband spectrum sharing scenario with limited rate feedback}
\label{s1}
\end{figure}

\begin{figure}[h]
\centering
\includegraphics[scale=0.45]{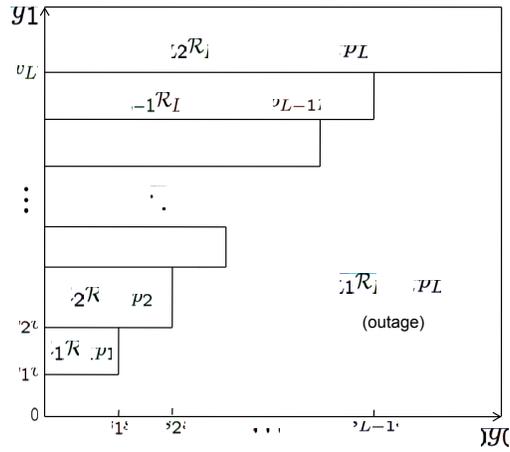}
\caption{The 'stepwise structure' of optimum quantization regions
for $\mu>0$ case} \label{r1}
\end{figure}

As shown in Fig.\ref{s1}, following our earlier work
\cite{he_dey_tcomm_11}, we assume that there is a central
controller termed as CR network manager who can obtain perfect
information of ${g}_0$ and $g_1$, from PU-RX at the PU base
station and SU-RX at the SU base station respectively, possibly
over fibre-optic links, and then forward some appropriately
quantized $(g_0, g_1)$ information to SU-TX through a finite-rate
feedback link. For further details on the justification of
resulting benefits due this assumption, see
\cite{he_dey_tcomm_11}.
%Note that the existence of such
%central controllers is also assumed quite commonly in literature
%on multi-cell MIMO or macro-diversity based systems with
%cooperative base stations in a primary network, where several base
%stations are assumed to be connected to a central controller via a
%backhaul link so that information about out-of-cell interference
%can be obtained resulting in higher
%capacity\cite{chen2007}\cite{ng2008}.
Under such a network modelling assumption, given B bits of
feedback, a power codebook ${\cal{P}}$$=\{ {p}_1, \dots, {p}_L\}$
of cardinality $L=2^B$, is designed offline purely on the basis of
the statistics of $g_0$ and $g_1$ information at the CR network
manager. This codebook is also known {\em a priori} by both SU-TX
and SU-RX for decoding purposes. Given a channel realization
$({g}_0, g_1)$, the CR network manager employs a deterministic
mapping from the current instantaneous $(g_0, g_1)$ information to
one of $L$ integer indices (let ${\cal {I}}(g_0, g_1)$ denote the
mapping, which partitions the vector space of $(g_0, g_1)$ into
$L$ regions $\cal{R}$$_1, \dots,$$\cal{R}$$_L$, defined as ${\cal
{I}}(g_0, g_1)=j,  ~~\text{if} ~~(g_0, g_1)\in
{\cal{R}}_j,~~j=1,\dots,L$),
%v\in[v_j, v_{j+1}) where $v_j, ~j=2,\dots,L$ represents the quantization point between
%$\cal{R}$$_{j-1}$ and ${\cal{R}}_j$, and $v_1=0$,
%$v_{L+1}=\infty$)
 and then sends the corresponding index $j={\cal
{I}}(g_0, g_1)$ to the SU-TX (and the SU-RX) via the feedback link. The SU-TX then uses
the associated power codebook element (e.g.,~if the feedback signal
is $j$, then $p_j$ will be used as the transmission power) to adapt
its transmission strategy.
\begin{remark}
Note that the CR network manager could be assumed to be located at the SU-BS for the current setup and in this case, the PU-BS simply has to cooperate
with the SU-BS by sending the real-time full CSI information of $g_0$. However, for future generalization of our work to a multi-cell cognitive network scenario, we
assume that the CR network manager is a separate entity, which can obtain information from multiple PU-BS and SU-BS if necessary.
\end{remark}
\indent\indent Define an indicator function $X_j, ~j=1,\dots,L,$
as $X_j=1$ if $\frac{1}{2}\log(1+g_1p_{j}) < r_0$, and $0$ otherwise.
Let $Pr ({ \cal {R}}_j)$, $E[\bullet|{\cal{R}}_j]$
represent $Pr ((g_0, g_1)\in { \cal {R}}_j)$ and $E[\bullet|(g_0, g_1)\in {\cal{R}}_j]$,
respectively. Then the SU outage probability minimization problem
 with limited feedback can be formulated as
\begin{align}
&\min_{p_j\geq 0,~ {\cal{R}}_j~ \forall j} ~\sum^L_{j=1}
E[X_j|{\cal{R}}_j]Pr ({
\cal {R}}_j) \nonumber\\
&~~s.t.~~~~~~~~~\sum^L_{j=1} E[p_j|{\cal{R}}_j]Pr ({
\cal {R}}_j)\leq P_{av},~~~~~~~\sum^L_{j=1} E[g_0p_j|{\cal{R}}_j]Pr ({
\cal {R}}_j)\leq Q_{av}.
 \label{uQ2}
\end{align}
Thus the key problem to solve here is the  joint optimization of   the channel partition regions and the power codebook such that the outage probability of SU is minimized under the above constraints.\\
%The Lagrangian of Problem  (\ref{optprob}) is expressed as,
%\begin{equation}
%L({\cal{P}},\lambda)=\sum_{j=1}^{L} E[x_j+\lambda{\textbf{P}_j}^{\sum}|{\cal{R}}_j]Pr ({ %\cal {R}}_j)-\lambda P_{av}
%\end{equation}
\indent The dual problem of (\ref{uQ2}) is expressed as,
$\max_{\lambda\geq 0,~\mu\geq 0}~ g(\lambda, \mu)-\lambda P_{av}-\mu Q_{av}$,
where $\lambda, \mu$ are the nonnegative Lagrange multipliers associated with the ATP and AIP constraints in Problem (\ref{uQ2}), and the Lagrange dual function $g(\lambda, \mu)$ is defined as
\begin{equation}
g(\lambda, \mu)=\min_{{p}_j\geq 0,~ {\cal{R}}_j,~\forall j } ~\sum_{j=1}^{L} E[X_j+(\lambda+\mu g_0)p_j|{\cal{R}}_j]Pr ({ \cal {R}}_j)
\label{udual}
\end{equation}
\indent The procedure we use to solve the above dual problem is:
\begin{enumerate}
\item [Step 1:] With fixed values of $\lambda$ and $\mu$, find the optimal solution (power codebook and quantization regions) for the Lagrange dual function (\ref{udual}). %there are two general methods: (1) one option is that
\item [Step 2:] Find the optimal $\lambda$ and $\mu$ by solving the dual problem using subgradient
search method, i.e, updating $\lambda,~ \mu$ until convergence using
\begin{align}
&\lambda^{l+1}= [\lambda^l-\alpha^l(P_{av}-\sum^L_{j=1}E[p_j|{\cal{R}}_j]Pr ({
\cal {R}}_j))]^+, \nonumber\\
&\mu^{l+1}= [\mu^l-\beta^l(Q_{av}-\sum^L_{j=1}E[g_0p_j|{\cal{R}}_j]Pr ({
\cal {R}}_j))]^+,
\label{subgra}
\end{align}
where $l$ is the iteration number, $\alpha^l$, $\beta^l$ are positive scalar step
sizes for the $l$-th iteration satisfying $\sum_{l=1}^{\infty} \alpha_l = \infty, \; \sum_{l=1}^{\infty} (\alpha_l)^2 < \infty$ and similarly for
$\beta_l$, and $[x]^+ = \max (x,0)$.
\end{enumerate}
\begin{remark}
A general method to solve  Step 1 is to employ a simulation-based
optimization algorithm called Simultaneous Perturbation Stochastic
Approximation (SPSA) algorithm (for a step-by-step guide to
implementation of SPSA, see \cite{spall98}), where one can use the
objective function of Problem (\ref{udual}) as the loss function
and the optimal power codebook elements for each channel partition
are obtained via a randomized stochastic gradient search
technique. Note that due to the presence of the indicator function
and no explicit expression being available for the outage
probability with quantized power allocation,
 we can't directly exploit the Generalized Lloyd Algorithm (GLA) with a Lagrangian distortion, as we used in \cite{he_dey_tcomm_11}, to solve  Problem (\ref{udual}). SPSA uses a simulation-based method to compute the loss function and then estimates the gradient from a number of loss function values
 computed by randomly perturbing the power codebook.  Note that SPSA results in a local minimum (similar to GLA), but is computationally highly complex and
 the convergence time is also quite long.
\end{remark}
\indent\indent Due to the high computational complexity of SPSA
and its long convergence time to solve Problem (\ref{udual}), we
will next derive a low-complexity approach for solving Problem
(\ref{udual}).  However,  due to the original problem (\ref{uQ2})
not being convex with respect to the power
codebook elements,  the optimal solution we can obtain is also locally optimal.\\
\indent Let ${\cal{P}}=\{p_1,\dots,p_L\}$, where
$p_1>\dots>p_L\geq 0$, and the corresponding channel partitioning
${\cal{R}}_1,\dots,{\cal{R}}_L$ denote an optimal solution to
Problem (\ref{udual}). Let $p({\cal I}(g_0, g_1))$ represent the
mapping from instantaneous $(g_0, g_1)$ information to the
allocated power level. We can then obtain the following result:
\begin{lemma}
\indent  Let $\{v_1,\dots,v_{L}\}$ denote the optimum
quantization thresholds on the $g_1$ axis ($0<v_1<\dots<v_{L}$)
and $\{s_1,\dots,s_{L-1}\}$ indicate the optimum quantization
thresholds on the $g_0$ axis ($0<s_1<\dots<s_{L-1}$). Then we have
$\forall j, j=1,\dots,L-1$, $p({\cal {I}}(g_0, g_1))=
p_j$, if $v_j\leq g_1 <v_{j+1}, 0\leq g_0 <s_{j}$ and 
$p_L$ otherwise, 
where $v_j=\frac{c}{p_j}, j=1,\dots,L$, and for $\forall j, j=1,\dots,L-1$, when $\mu> 0,~s_j=\frac{1}{\mu(p_j-p_L)}-\frac{\lambda}{\mu}$, while when $\mu= 0,~s_j=\infty$, then condition $0\leq g_0 <s_{j}$ boils down to $\lambda<\frac{1}{p_j-p_L}$. The region ${\cal{R}}_L$ includes two parts : the set ${\cal{R}}_{L1}=\{(g_0, g_1):v_j\leq g_1 <v_{j+1}, g_0 \geq s_{j}, \forall j=0,\dots,L-1\}$ with $s_0=0, v_0=0$ and the set ${\cal{R}}_{L2}=\{(g_0, g_1):g_1 \geq v_L, g_0\geq 0\}$. The entire set ${\cal{R}}_{L1}$ is in outage.
 \label{ul1}
\end{lemma}
\indent\indent\textit{Proof:} See Appendix \ref{app0}.
\begin{remark}
When $\mu>0$, which implies that the AIP constraint is active, from
Lemma \ref{ul1}, the optimum partition regions possess a
stepwise structure, as shown in Fig.\ref{r1}. When $\mu=0$, i.e, the AIP
constraint is inactive and only ATP constraint is active (we must
have $\lambda>0$), Problem (\ref{uQ2}) becomes a scalar
quantization problem involving quantizing $g_1$ only, and Lemma
\ref{ul1} reduces to : $p({\cal I}(g_1))=
p_j$, if $v_j\leq g_1 <v_{j+1}, \forall j, j=1,\dots,L-1$, and 
$p_L$ otherwise,
where $\lambda<\frac{1}{p_j-p_L}, \forall j=1,2,\ldots, L-1$ and the two sub-regions of ${\cal{R}}_L$ become ${\cal{R}}_{L1}=\{g_1:0\leq g_1<v_1\}$ and ${\cal{R}}_{L2}=\{g_1:g_1 \geq v_L\}$, and ${\cal{R}}_{L1}$ is in outage. Note that in this case we must have $Q_{av}\geq P_{av}$, due to $Q_{av}\geq \sum^L_{j=1} E[g_0p_j|{\cal{R}}_j]Pr ({\cal {R}}_j)=\sum^L_{j=1}E[p_j|{\cal{R}}_j]Pr ({\cal {R}}_j)=P_{av}$, where the last equality follows from the fact the
$E[g_0 |{\cal R}_j] = E[g_0] =1$ since ${\cal {R}}_j$ is formed purely based on the values of $g_1$, which is  independent of $g_0$.
Note also that one can easily prove the converse, that when $Q_{av} \geq P_{av}$, one must have $\mu=0$.
\end{remark}
 %when $\lambda>0$,$p_L=v_L=(\frac{1}{\lambda}-\frac{1}{g_1})^+$ which is not a
%constant, this is a bit unusual in doing normal quantization,
%however, since both SU-TX and PU-TX have perfect knowledge of
%$g_1$, we can treat it as a 'constant'; where ${\tilde{f}_1}(v_1)=0$ and
\indent\indent From Lemma \ref{ul1}, (due to the fading channels being independently exponentially distributed with unity mean) Problem (\ref{uQ2}) becomes,
\begin{align}
&\min_{p_j\geq 0,~\forall j} ~P^L_{out}=1-e^{-v_1}+\sum^{L-1}_{j=1}(e^{-v_j}-e^{-v_{j+1}})e^{-s_j}\ \nonumber\\
&~~s.t.~~p_L+\sum^{L-1}_{j=1}(p_j-p_L)(e^{-v_j}-e^{-v_{j+1}})(1-e^{-s_j}) \leq P_{av} \nonumber\\
&~~~~~~~~p_L+\sum^{L-1}_{j=1}(p_j-p_L)(e^{-v_j}-e^{-v_{j+1}})(1-e^{-s_j}(1+s_j)) \leq Q_{av}
 \label{uQ3}
\end{align}
where $P^L_{out}$ denotes the outage probability with $B=\log_2 L$
bits feedback QPA, $v_j=\frac{c}{p_j}, j=1,\dots,L$ and for
$\forall j, j=1,\dots,L-1$, when $\mu>
0,~s_j=\frac{1}{\mu(p_j-p_L)}-\frac{\lambda}{\mu}$, whereas when
$\mu= 0,~s_j=\infty$. Although the above optimization problem may
be verified to be non-convex, we can employ the KKT necessary
conditions to find a local minimum for Problem (\ref{uQ3}). Taking
the partial derivative of first order of the Lagrangian
 of Problem (\ref{uQ3}) over $p_j, j=1,\dots,L-1$, and
setting it to zero, we can obtain
\begin{align}
(e^{-v_j}-e^{-v_{j+1}})[\lambda(1-e^{-s_j})+\mu(1-e^{-s_j}(1+s_j))]=e^{-v_j}\frac{c}{p^2_j}[\hat{f}(p_{j-1})-\hat{f}(p_j)], ~~1\leq j \leq L-1;
\label{ukkt1}
\end{align}
where $\hat{f}(p_0)=1$ and $\hat{f}(p_j)=(p_j-p_L)(\lambda+\mu(1-e^{-s_{j}})),~1\leq j\leq L-1$.
(\ref{ukkt1}) also can be rewritten as $j=1,\dots,L-1$,
\begin{align}
p_{j+1}=\frac{c}{v_j-\ln (1-\frac{\frac{c}{p^2_j}[\hat{f}(p_{j-1})-\hat{f}(p_j)]}{\lambda(1-e^{-s_j})+\mu(1-e^{-s_j}(1+s_j))})},
\label{ukkt4}
\end{align}
Equating the partial derivative of the Lagrangian
function of Problem (\ref{uQ3}) over $p_L$  to zero gives,
\begin{align}
\sum^{L-1}_{j=1}(e^{-v_j}-e^{-v_{j+1}})[\lambda(1-e^{-s_j})+\mu(1-e^{-s_j}(1+s_j))]+e^{-v_L}\frac{c}{p^2_L}\hat{f}(p_{L-1})=\lambda+\mu.
\label{ukkt2}
\end{align}
Optimal values of $\lambda$ and $\mu$ can be determined by solving
\begin{align}
&\lambda[p_L+\sum^{L-1}_{j=1}(p_j-p_L)(e^{-v_j}-e^{-v_{j+1}})(1-e^{-s_j})-P_{av}]=0\nonumber\\
&\mu[p_L+\sum^{L-1}_{j=1}(p_j-p_L)(e^{-v_j}-e^{-v_{j+1}})(1-e^{-s_j}(1+s_j))-Q_{av}]=0
\label{ukkt3}
\end{align}
Thus, for fixed values $\lambda$ and $\mu$, we need to solve the $L$ equations given by (\ref{ukkt4}), (\ref{ukkt2}) to obtain the power codebook. Given $p_1$ and $p_L$, from (\ref{ukkt4}) we can
successively compute $p_{2},\dots,p_{L-1}$, and then we can jointly solve the equation (\ref{ukkt4}) with $j=L-1$ and equation (\ref{ukkt2}) numerically for $p_1$ and $p_L$.
The optimal value of $\lambda$ and $\mu$ can be obtained by
solving (\ref{ukkt3}) with a subgradient method, i,e. by updating
$\lambda$ and $\mu$ until convergence using (\ref{subgra}).
One can thus repeat the above two
steps (i.e, given $\lambda$ and $\mu$ find the optimal power levels, and then using the resulting optimal power levels update $\lambda$ and $\mu$)
iteratively until a satisfactory convergence criterion is met. This procedure can be formally summarized as:
\begin{enumerate}
\item[a)] First, if $P_{av}\leq Q_{av}$, we must have $\mu=0,~\lambda>0$. Starting with an arbitrary positive initial value for $\lambda$, solve (\ref{ukkt1}), (\ref{ukkt2}) to obtain a power codebook $\{p_1,\dots,p_L\}$, and then use this codebook to update $\lambda$ by (\ref{subgra}).
Repeat these steps until convergence and the final codebook will be an optimal power codebook for Problem (\ref{uQ3}).
\item[b)] If $P_{av}> Q_{av}$, we must have $\mu>0$ by contradiction (since if $\mu=0$, we must have
$P_{av} \leq Q_{av}$). Let $\lambda=0$, then solving KKT conditions gives an optimal value of $\mu$ and corresponding power codebook $\{p_1,\dots,p_L\}$. With this codebook, if $\sum^L_{j=1} E[p_j|{\cal{R}}_j]Pr ({
\cal {R}}_j)\leq P_{av}$, then it is an optimal power codebook for Problem (\ref{uQ3}). Otherwise we must have $\lambda>0$ too, in which case, starting with  arbitrary  positive initial values for $\lambda$ and $\mu$,  obtain the corresponding power codebook $\{p_1,\dots,p_L\}$, and then update $\lambda$ and $\mu$ by (\ref{subgra}). Repeat these steps until convergence and the final codebook will be an optimal power codebook for Problem (\ref{uQ3}).
\end{enumerate}
%\begin{remark}
%To solve Problem (\ref{uQ3}), we also can exploit various
%optimization softwares, like "1stOpt" software (also known as
%"Auto2Fit"), which uses the Levenberg-Marquardt + Universal Global
%Optimization method to find the solution, and does not require
%good estimates of initial values for the optimization variables,
%but not fit for dealing with large number of variables.
%\end{remark}
\subsection{Suboptimal QPA Algorithm}
When the number of feedback bits $B$ (or alternatively, $L$) goes to infinity, we can obtain the following Lemma that allows
us to obtain a suboptimal but  computationally efficient
quantized power allocation algorithm for large but finite $L$.
\begin{lemma}
$\lim_{L\rightarrow\infty} p_L=0$
\label{ul2}
\end{lemma}
\indent\indent\indent\textit{Proof:} See appendix \ref{app4}.
\begin{remark}
Lemma \ref{ul2} shows that regardless of whether $\mu>0$ or $\mu=0$, with high rate quantization, the power
level for the last region
${\cal{R_L}}$ approaches zero, which also implies the following as $L\rightarrow\infty$:\\
 1) The non-outage part of ${\cal{R_L}}$, given by ${\cal{R_L}}_2$,
disappears gradually.  In other words, ${\cal{R_L}}\rightarrow
{\cal{R_L}}_1$. Thus, when $L\rightarrow\infty$, ${\cal{R_L}}$
becomes the outage region with zero power assigned to it. \\
2) When $\mu> 0$, the
quantization thresholds on the $g_0$ axis $s_j\rightarrow s_j'$ (where $s_j'=\frac{1}{\mu p_j}-\frac{\lambda}{\mu}$), which gives
$v_j=c\lambda+c\mu s'_j$, and it means all the points given by coordinates $(s'_j, v_j)$ lie on the line of $g_1=c\lambda+c\mu g_0$. Therefore, as $L\rightarrow\infty$, the stepwise shape of the structure in $\mu> 0$ case (i.e, the boundary between non-outage and outage regions) approaches the straight line $g_1=c\lambda+c\mu g_0$, which is consistent with the full CSI-based power allocation result in \cite{Kang09}.
\end{remark}
\indent\indent Thus, when $L$ is large, applying Lemma \ref{ul2}
(i.e, $p_L\rightarrow0$) to Problem (\ref{uQ3}), the above $L$ KKT
conditions (\ref{ukkt1}) and (\ref{ukkt2}) can be simplified into
$L-1$ equations:
\begin{align}
&(e^{-v_j}-e^{-v_{j+1}})[\lambda(1-e^{-s'_j})+\mu(1-e^{-s'_j}(1+s'_j))]\nonumber\\
&=e^{-v_j}\frac{c}{p^2_j}[p_{j-1}(\lambda+\mu(1-e^{-s'_{j-1}}))-p_j(\lambda+\mu(1-e^{-s'_{j}}))], \; \forall j, j=1,\dots,L-1
\label{ukkt6}
\end{align}
where when $\mu> 0$, the quantization thresholds on the $g_0$ axis
are given by $s'_j=\frac{1}{\mu p_j}-\frac{\lambda}{\mu}$, $s'_0=0$, and
$p_0=\frac{1}{\lambda+\mu s'_0}$, while when $\mu= 0,~s'_j=\infty,
s'_0=0$, and $p_0=\frac{1}{\lambda}$. (\ref{ukkt6}) can be also
written as
\begin{align}
&p_{j+1}=\frac{c}{v_j-\ln
(1-\frac{\frac{c}{p^2_j}[p_{j-1}(\lambda+\mu(1-e^{-s'_{j-1}}))-p_j(\lambda+\mu(1-e^{-s'_{j}}))]}{\lambda(1-e^{-s'_j})+\mu(1-e^{-s'_j}(1+s'_j))})},
~~~j=1,\dots,L-2;\nonumber\\
&\frac{\lambda(1-e^{-s'_{L-1}})+\mu(1-e^{-s'_{L-1}}(1+s'_{L-1}))}{\frac{c}{p^2_{L-1}}[p_{L-2}(\lambda+\mu(1-e^{-s'_{L-2}}))-p_{L-1}(\lambda+\mu(1-e^{-s'_{{L-1}}}))]}=1
\label{ukkt5}
\end{align}
Thus, for given values of $\lambda$ and $\mu$, starting with a specific value of
$p_1$, we can successively compute $p_2,\dots,p_{L-1}$ using
the first equation of (\ref{ukkt5}) (recall that $v_j = \frac{c}{p_j}$). Then the second  equation in (\ref{ukkt5})
becomes an equation in $p_1$ only, which can be
solved easily using a suitable nonlinear equation solver.  We
call this suboptimal QPA algorithm as 'Zero Power in Outage Region
Approximation'(ZPiORA), which is applicable to the case
of a large number of feedback bits, where the exact definition of ``large" will be dependent on the system parameters. Through simulation studies, we will illustrate that for our choice of system parameters, ZPiORA performs well even for as low as $B=2$ bits of feedback. \\
{\bf Alternative suboptimal algorithms}:
For comparison purposes, we also propose two alternative  suboptimal algorithms described below:
\begin{enumerate}
\item[(1)] The first suboptimal algorithm is based on an equal average
power per (quantized) region (EPPR) approximation algorithm, proposed in
\cite{icc} in a non-cognitive or typical primary network setting for an outage minimization problem
with only an ATP constraint. More specifically, by applying the mean value theorem (similar to \cite{icc}) into the KKT conditions (\ref{ukkt1}) with $j=2,\dots,L-1$, we can easily obtain that $p_j(e^{-v_j}-e^{-v_{j+1}})[\lambda(1-e^{-s_j})+\mu(1-e^{-s_j}(1+s_j))]
\approx p_{j-1}(e^{-v_{j-1}}-e^{-v_{j}})[\lambda(1-e^{-s_{j-1}})+\mu(1-e^{-s_{j-1}}(1+s_{j-1}))], ~~j=2,\dots,L-2$. Adding the two
equations of (\ref{ukkt3}) together and applying (\ref{ukkt2}), we have $\sum^{L-1}_{j=1} p_j (e^{-v_j}-e^{-v_{j+1}})[\lambda (1-e^{-s_j})+\mu (1-e^{-s_j}(1+s_j))]=\lambda P_{av}+\mu Q_{av}-e^{-v_L}\frac{c}{p_L}(p_{L-1}-p_{L})(\lambda+\mu(1-e^{-s_{L-1}}))$. Since $e^{-v_L}\frac{c}{p_L}(p_{L-1}-p_{L})(\lambda+\mu(1-e^{-s_{L-1}}))$ can be approximated as $p_{L-1}(e^{-v_{L-1}}-e^{-v_{L}})[\lambda(1-e^{-s_{L-1}})+\mu(1-e^{-s_{L-1}}(1+s_{L-1}))]$ by using the mean value theorem, we can obtain the following $L$ (approximate) equations, namely $p_j(e^{-v_j}-e^{-v_{j+1}})[\lambda(1-e^{-s_j})+\mu(1-e^{-s_j}(1+s_j))] \approx \frac{\lambda P_{av}+\mu Q_{av}}{L},~j=1,\dots,L-1$ and $p_L(1-\sum^{L-1}_{j=1} (e^{-v_j}-e^{-v_{j+1}})[\lambda(1-e^{-s_j})+\mu(1-e^{-s_j}(1+s_j))]) \approx \frac{\lambda P_{av}+\mu Q_{av}}{L}$. Then one can jointly solve the above $L$ equations and  two other equations ((\ref{ukkt1}) with $j=1$ and (\ref{ukkt2})) for $\lambda, \mu, p_j, \forall j=1,\dots,L$. We call this suboptimal algorithm as the ``Modified EPPR (MEPPR)" approximation algorithm. Obviously, ZPiORA is computationally much simpler than this method, especially when $\mu>0$. Furthermore, from simulations, when $P_{av}$ or ${Q_{av}}$ is small, the performance of ZPiORA is
always better than MEPPR. It is seen however that when both $P_{av}$ and
$Q_{av}$ are large, for a small number of feedback bits, MEPPR may outperform ZPiORA, whereas with a sufficiently large number of feedback
bits, ZPiORA is a more accurate approximation due to Lemma $\ref{ul2}$ (when $L$ is large, $p_L$ approaches zero, whereas MEPPR  has $p_L>0$ $\forall L$). See Section V for more details.
Note that, an equal probability per region
(excluding the outage region) approximation algorithm employed in \cite{KRZhang09} for scalar quantization can not be applied to our case (vector quantization), since it will increase the computational complexity even further.
\item[(2)] The second algorithm is based on GLA with a sigmoid function approximation (GLASFA) method proposed by \cite{weiyu2009}, where the sigmoid function is used to approximate the indicator function in the Lagrange dual function (\ref{udual}). More specifically, given a random initial power codebook, we use the nearest neighbor condition of Lloyd's algorithm with a Lagrangian distortion $d((g_0, g_1),j)=X_j+(\lambda+\mu g_0)p_j$ to generate the optimal partition regions \cite{Gray89} given by, ${\cal{R}}_j=\{(g_0,g_1): X_j+(\lambda+\mu g_0)p_j\leq X_i+(\lambda+\mu g_0)p_i, \forall i\not=j\}$,  $i,j=1,\dots,L$.
We then use the resulting optimal partition regions to update the power codebook by
$p_j\approx \text{argmin}_{p_j\geq 0} E[\sigma(k(\frac{1}{2}\log(1+g_1p_{j})-r_0))+(\lambda+\mu g_0)p_j|R_{j}]Pr(R_{j})$ for $ j=1,\dots,L$, where we use the approximation $X_j\approx \sigma(k(\frac{1}{2}\log(1+g_1p_{j})-r_0))$,  $\sigma(x)=\frac{1}{1+e^x}$ being the sigmoid function where the coefficient $k$ controls the sharpness of the approximation (for detailed guidelines on choosing $k$ see \cite{weiyu2009}). The above two steps of GLA are repeated until convergence.
%Therefore, with given power codebook and resulting quantization regions, we can numerically calculate the loss function, which is then used for updating the new power codebook. We repeatedly apply the step 2 to step 5 of SPSA in \cite{spall98} until the resulting outage probability converges within a pre-specified accuracy (i.e, the step 6 of SPSA in \cite{spall98} is satisfied). %(2) we also can exploit the modified Generalized Lloyd Algorithm (GLA) method as we used in \cite{paper2010}, i.e, a). given a power codebook, the optimal partitions is given by equation (\ref{n1}), and then,  b). with designed optimal partitions, the new optimal power codebook is found by solving for
%\begin{equation}
%p_j=\text{argmin}_{p_j\geq 0} E[X_j+(\lambda+\mu g_0)p_j|{\cal{R}}_j]Pr ({ \cal {R}}_j). \forall j
%\end{equation}
%Due to hard to find the closed-form of $E[X_j|{\cal{R}}_j]$, to solve above Problem we need to use an approximation $X_j\approx\sigmoid(k(x-c))$ as \cite[weiyu2009] did.
%These two steps a) and b) are repeated until convergence.
Numerical results illustrate that ZPiORA significantly outperforms this suboptimal method. See Section V for more details.
\end{enumerate}
\section{Asymptotic outage behaviour of QPA under high resolution quantization}\label{us3}
\indent In this section, we derive a number of asymptotic expressions for the SU outage probability when the
number of feedback bits approaches infinity. To this end, we  first derive some useful properties regarding the quantizer structure at high rate
quantization:
\begin{lemma}
As the number of quantization regions $L\rightarrow\infty$, we can obtain the following result:
with $\mu>0$, the optimum quantization thresholds on the $g_0$ axis satisfy \\
$s'_{1}-s'_{0}\approx s'_{2}-s'_{1}\approx \dots \approx
s'_{L-1}-s'_{L-2}$,
where $s'_j=\frac{1}{\mu p_j}-\frac{\lambda}{\mu}, j=1,\dots,L-1$ and $s'_0=0$.
With $\mu=0$, the optimum quantization thresholds on the $g_1$ axis satisfy
$\frac{v_{1}}{v_{0}} \approx \frac{v_{2}}{v_{1}} \dots \approx
\frac{v_{L-1}}{v_{L-2}}$,
where $v_j=\frac{c}{p_j}, j=1,\dots,L-1$ and here $v_0=c\lambda$.
\label{ul3}
\end{lemma}
\indent\indent\indent\textit{Proof:} See Appendix \ref{app1}.
\begin{lemma}
In the high rate quantization regime, as  $L\rightarrow \infty$, we have
\begin{align}
\sum^{L-1}_{j=1} (e^{-v_j}-e^{-v_{j+1}})[\lambda(1-e^{-s'_j})+\mu(1-e^{-s'_j}(1+s'_j))]\approx \frac{\lambda P_{av}+\mu Q_{av}}{L-1}
\sum^{L-1}_{j=1} \frac{1}{p_j}.
\label{uk20}
\end{align}
where when $\mu> 0$, $s'_j=\frac{1}{\mu p_j}-\frac{\lambda}{\mu}$, whereas when $\mu= 0,~s'_j=\infty$, and (\ref{uk20}) simplifies to
$ce^{-v_1}\approx \frac{P_{av}}{L-1}\sum^{L-1}_{j=1} v_j$ with $v_j=\frac{c}{p_j}$.
\label{ul4}
\end{lemma}
\indent\indent\indent\textit{Proof:} See Appendix \ref{app2}.\\
\indent With Lemma \ref{ul3} and Lemma \ref{ul4}, the main result
of this section can be obtained in the following Theorem.
\begin{theorem}
The asymptotic SU outage probability  for a large number of feedback bits is given as,\\
$P^L_{out}\approx
1-e^{-c\lambda_f^*}[1-(1-e^{-\frac{a}{L}})\frac{1-e^{-a(1+\frac{1}{c\mu_f^*})}}{1-e^{-\frac{a(1+\frac{1}{c\mu_f^*})}{L}}}]$
(for $\mu > 0$)
where $a$ is a constant satisfying
\begin{align}
&(\lambda_f^* P_{av}+\mu_f^* Q_{av})(\lambda_f^*+\frac{a}{2c})e^{c\lambda_f^*}\nonumber\\
&\approx [(\lambda_f^*+\mu_f^*)(1-\frac{c\mu_f^*}{1+c\mu_f^*}(1-e^{-a(1+\frac{1}{c\mu_f^*})}))-\frac{c (\mu_f^*)^2}{(1+c\mu_f^*)^2
}(1-e^{-a(1+\frac{1}{c\mu_f^*})}(1+a(1+\frac{1}{c\mu_f^*})))].
\end{align}
We also have
$\lim_{L\rightarrow \infty} P^L_{out}=
1-e^{-c\lambda_f^*}[1-\frac{1-e^{-a(1+\frac{1}{c\mu_f^*})}}{1+\frac{1}{c\mu_f^*}}]$. For $\mu=0$,
$P^L_{out}\approx 1-e^{-c\lambda_f^* (1+\frac{\beta}{L})}$,
where $\beta$ is a constant given by
$ e^{-c\lambda_f^*}\approx {\lambda_f^*
P_{av}}\frac{e^{\beta}-1}{\beta}$.
In this case we also have
$\lim_{L\rightarrow \infty} P^L_{out}= 1-e^{-c\lambda_f^*}$.
\label{ut1}
\end{theorem}
\indent\indent\indent\textit{Proof:} See Appendix \ref{app3}.
\section{Numerical Results}\label{us4}
\begin{figure}[h]
\centering
\includegraphics[scale=0.5]{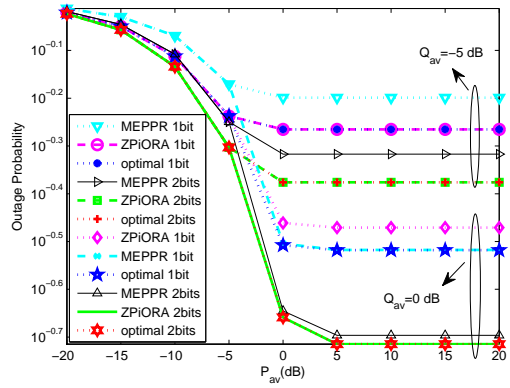}
\caption{Outage probability performance comparison between ZPiORA, MEPPR and optimal QPA}
\label{f1}
\end{figure}

\begin{figure}[h]
\centering
\includegraphics[scale=0.5]{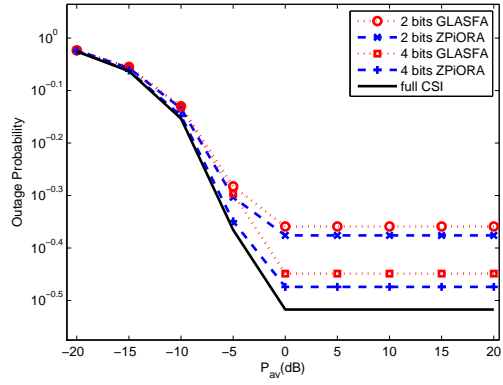}
\caption{Outage probability performance comparison between ZPiORA and other possible suboptimal algorithm : GLASFA}
\label{f5}
\end{figure}

\begin{figure}[h]
\centering
\includegraphics[scale=0.5]{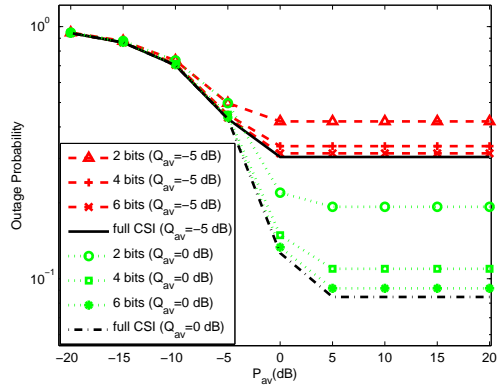}
\caption{Effect of increasing feedback bits on outage performance of SU}
\label{f2}
\end{figure}

\begin{figure}[h]
\centering
\includegraphics[scale=0.5]{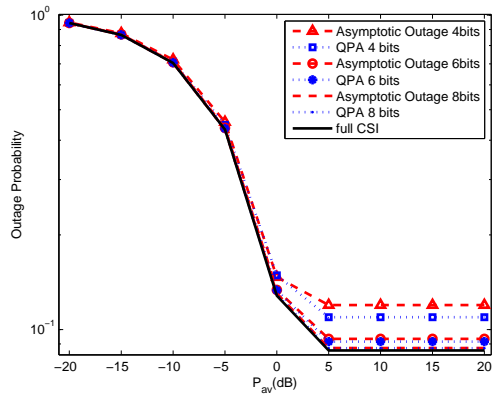}
\caption{Comparison between asymptotic outage performance and QPA performance with $Q_{av}=0 dB$ }
\label{f3}
\end{figure}

\begin{figure}[h]
\centering
\includegraphics[scale=0.5]{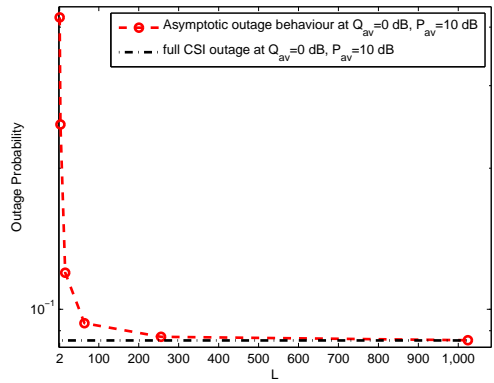}
\caption{Asymptotic outage behaviour versus the number of
quantization level $L$}
\label{f4}
\end{figure}

In this section, we will examine the outage probability
performance of the SU in a narrowband spectrum sharing system
with the proposed power allocation strategies via numerical
simulations. All the channels involved are assumed to be independent and undergo
identical Rayleigh fading, i.e, channel power gain $g_{0}$ and
$g_1$ are independent and identically
 exponentially distributed with unity mean.  The required transmission rate is taken to be $r_0=0.25$ nats per channel use.\\
\indent Fig. \ref{f1} displays the SU outage probability performance of the suboptimal algorithm
ZPiORA versus $P_{av}$ with feedback bits $B=\{1, 2\}$, under $Q_{av}=-5$ dB and $Q_{av}=0$ dB respectively, and compares these results with the corresponding outage performance of the suboptimal method MEPPR and the optimal QPA. As observed from Fig. \ref{f1}, when $Q_{av}=-5$ dB, with $B$ fixed, the outage performances of ZPiORA and corresponding optimal QPA  almost overlap with each other. When $Q_{av}=0$ dB and $P_{av}\leq -5$ dB, with the same number of feedback bits, the outage performances of these two methods are still  indistinguishable; and with $P_{av}>-5$ dB, the outage performance gap between ZPiORA and corresponding optimal QPA is decreasing with increasing B. For example, with 1 bit feedback, at $P_{av}=10$ dB, the outage gap between ZPiORA and optimal QPA is $0.0347$, but with 2 bits of feedback, the outage performance of these two methods are very close to each other, which agrees with Lemma \ref{ul2} that ZPiORA is a near-optimal algorithm for large number of feedback bits. Now we look at the performance comparison between ZPiORA and MEPPR. As illustrated in Fig. \ref{f1}, when $P_{av}$ or ${Q_{av}}$ is small, with B bits feedback, the performance of ZPiORA is
always better than MEPPR. This is attributed to the fact that when $P_{av}$ or ${Q_{av}}$ is small, it can be easily verified that $p_L$ is close to zero, but MEPPR always uses $p_L>0$. However, when both $P_{av}$ and
$Q_{av}$ are large (e.g. $P_{av}\geq 0$ dB and $Q_{av}=0$ dB), for 1 bit feedback case, MEPPR outperforms ZPiORA and performs very close to the optimal QPA, whereas with a sufficiently large number of feedback
bits (in fact, with more than just 2 bits of feedback), ZPiORA is a more accurate approximation due to Lemma $\ref{ul2}$.
These results confirm the ZPiORA is a better option for a large number of feedback bits, not to mention that ZPiORA is much simpler to implement than MEPPR.
\\\indent In addition, Fig. \ref{f5} compares the outage performance of ZPiORA with another suboptimal method (GLASFA) with $Q_{av}= -5$ dB. We can easily observe that  with a fixed number of feedback bits (2 bits or 4 bits), ZPiORA always outperforms GLASFA. And ZPiORA is also substantially faster than GLASFA. For example, with fixed $\lambda$ and
$\mu$ and 4 bits of feedback ($Q_{av}= -5$dB, $P_{av}= 10$ dB), when
implemented in MATLAB (version 7.11.0.584 (R2010b)) on a AMD Quad-Core processor (CPU P940 with a clock speed of  $1.70$ GHz
and a memory of 4 GB), it was seen that GLASFA (with 100,000 training
samples, starting $k=20$ and increasing it by a factor of 1.5 at each
step which finally converged at about $k=768.8672$) took approximately 299.442522 seconds (different initial guesses of the power codebook may result in  different convergence time).
In comparison, ZPiORA took only 0.006237 seconds to achieve comparable levels
of accuracy. These results further confirm the efficiency of ZPiORA.
\\
\indent Fig. \ref{f2} illustrates the outage performance
of SU with optimal QPA strategy versus $P_{av}$ with feedback bits $B=\{2,4,6\}$, under $Q_{av}=-5$ dB and $Q_{av}=0$ dB respectively, and studies the effect of increasing the number of feedback bits on the outage performance. For comparison, we
also plot the corresponding SU outage performance with full CSI case.  Since ZPiORA is an  efficient suboptimal method for large number of feedback bits,  we employ ZPiORA to obtain the outage performance instead of using optimal QPA for $B=6$ bits. First, it can be easily observed that all the outage curves decrease gradually as $P_{av}$ increases until $P_{av}$ reaches a
certain threshold,  when the outage probability attains a floor. This is due to the fact that in the high $P_{av}$ regime,
the AIP constraint dominates. For a fixed number of
feedback bits, the higher $Q_{av}$ is, the smaller the resultant outage probability is, since higher $Q_{av}$ means PU can tolerate more interference.
 Fig. \ref{f2} also illustrates  that for fixed
$Q_{av}$, introducing one extra bit of feedback substantially reduces the outage
gap between QPA and the perfect CSI case.
 To be specific, for $Q_{av}=0$ dB and $P_{av}=10$ dB, with
$2$ bits, $4$ bits and $6$ bits of feedback, the outage gaps with the full CSI case are approximately
$0.1083$, $0.0249$ and $0.006979 $ respectively. And for any $Q_{av}$,
only 6 bits of feedback seem to  result in an SU outage performance very
close to that with full CSI case.\\
\indent Figure \ref{f3} compares the asymptotic outage performance derived in Theorem \ref{ut1} and the optimal QPA performance  $B=\{4,6,8\}$ under
$Q_{av}= 0 $ dB.   It is clearly observed that increasing number of feedback bits substantially shrinks the outage performance gap between the asymptotic outage approximation and the corresponding optimal QPA performance.  For instance, with 4, 6, 8 bits of feedback at $P_{av}=10$ dB, the outage gap between the asymptotic outage approximation and the corresponding optimal QPA  are around $0.0325$,  $0.00618$, $0.000168$ respectively. These results confirm that the derived asymptotic outage expressions in Theorem \ref{ut1} are highly accurate for $B \geq 8$ bits of feedback.
In addition, Figure \ref{f4} depicts the asymptotic outage probability behavior of
SU versus the number of quantization level $L$ at $Q_{av}=0$ dB, $P_{av}=10$ dB, and compares the result with the corresponding full CSI performance. It can be seen from
Figure \ref{f4} that the outage decreases as the
number of quantization level $L$ increases, however, as $L$
increases beyond a certain number ($L\geq 2^8$, i.e, $B \geq 8$ bits), the
outage probability curve starts to saturate and approaches the full CSI performance. This further confirms that
only a small number of feedback bits is enough to
obtain an outage performance close to the perfect CSI-based performance.
\section{Conclusions and extensions}\label{us5}
In this paper, we designed optimal power allocation algorithms for secondary outage probability minimization with
quantized CSI information for a narrowband spectrum sharing
cognitive radio framework  under an ATP constraint at SU-TX and an
AIP constraint at PU-RX. We prove that the optimal channel partition structure
has a ``stepwise" pattern  based on which
an efficient optimal power codebook design algorithm is
provided. In the case  of a large number of feedback bits, we derive a novel low-complexity suboptimal algorithm ZPiORA which is seen to outperform
alternative suboptimal algorithms based on approximations used in the existing literature.  We also derive
explicit expressions for asymptotic behavior of the SU outage
probability for a large number of feedback bits. Although the presented optimal power codebook design methods result in locally optimal
solutions (due to the non-convexity of the quantized power allocation problem), numerical results illustrate that only 6 bits of feedback
result in SU outage performance very close to that obtained with full CSI at the SU transmitter.
Future work will involve extending the results to more complex
wideband spectrum sharing scenario along with consideration of other types of interference constraints at the PU receiver.
\begin{appendix}
\subsection{Proof of Lemma \ref{ul1}:}\label{app0}
\indent We use an analysis method similar to \cite{khoj08} to prove our problem's optimal quantizer structure. Let ${\cal{P}}=\{p_1,\dots,p_L\}$, where $p_1>\dots>p_L\geq 0$, and the corresponding channel partitioning ${\cal{R}}=\{{\cal{R}}_1,\dots,{\cal{R}}_L\}$ denote the optimal solution to the optimization problem (\ref{uQ2}), and $p(g_0, g_1)=p_j,~~ \text{if}~~ (g_0, g_1)\in {\cal{R}}_j$.\\
\indent Let ${\cal{R}}^*_j=\{(g_0,g_1): v_j\leq g_1 <v_{j+1}, 0\leq g_0 <s_{j}\},~ j=1,\dots,L-1$ and ${\cal{R}}^*_L={\cal{R}}^*_{L1}\cup{\cal{R}}^*_{L2}=\{(g_0,g_1): v_j\leq g_1 <v_{j+1},~g_0\geq s_{j}, \forall j=0,1,\dots,L-1\}\cup\{(g_0,g_1): g_1\geq v_{L}, g_0\geq 0\}$, where $s_0=0$ and $v_0=0$. We assume
that the set ${\cal{R}}^*_j \setminus {\cal{R}}_j$ is a non-empty set, where $\setminus$ is the set subtraction operation (i.e, if $(g_0, g_1)\in {\cal{R}}^*_j \setminus {\cal{R}}_j$, then $(g_0, g_1)\in{\cal{R}}^*_j$ but $(g_0, g_1)\notin{\cal{R}}_j$). Then, the set ${\cal{R}}^*_j \setminus {\cal{R}}_j$ can be partitioned into two subsets $S^-_j=({\cal{R}}^*_j \setminus {\cal{R}}_j)\cap(\cup^{j-1}_{k=1} {\cal{R}}_k) $ and $S^+_j=({\cal{R}}^*_j \setminus {\cal{R}}_j)\cap(\cup^{L}_{k=j+1}{\cal{R}}_k)$. In what follows, we denote the empty set by $\emptyset$.\\
{\bf (1)}: We will show that $S^-_j=\emptyset, ~\forall j=1,\dots,L$. \\
{\em (a)}: When $j=1$, it is  obvious that $S^-_1=\emptyset$. When
$1<j<L$, if $S^-_j\not=\emptyset$, then we can always reassign the
set $S^-_j$ into region ${\cal{R}}_j$ without changing the overall
outage probability. This is due to the fact that within the set
$S^-_j\in {\cal{R}}^*_j$, we have $v_j\leq g_1 <v_{j+1}$ resulting
in $\frac{1}{2}\log(1+g_1p_j)\geq r_0$, and the power level in
$(\cup^{j-1}_{k=1} {\cal{R}}_k)$ satisfies $p_k>p_j$. Thus $S^-_j$
is never in outage. However, the new assignment can achieve a
lower Lagrange dual function (LDF) in (\ref{udual}), due to
$g'(\lambda, \mu)-g(\lambda, \mu)=E[(\lambda+\mu
g_0)(p_j-p_k)|S^-_j]Pr (S^-_j)<0$, where $g'(\lambda, \mu)$
denotes the LDF with the new assignment, which contradicts the
optimality of the solution ${\cal{P}}, {\cal{R}}$. \\
{\em (b)} When
$j=L$, if $S^-_L\not=\emptyset$, we can again reassign the set
$S^-_L$ into region ${\cal{R}}_L$.  1) If some part of $S^-_L$ is in the set
$\{(g_0,g_1): 0\leq g_1 <v_1,~g_0\geq 0\}$ of ${\cal{R}}^*_{L1}$, we have $\frac{1}{2}\log(1+g_1p_1)< r_0$, which
implies that this part of $S^-_L$ is always in outage. Therefore,
this reassignment for this part of $S^-_L$ will not change
the outage probability but will decrease the LDF due to the power
level $p_L$ in ${\cal{R}}_L$ is the lowest. 2) For any $j$ ($j=1,\dots,L-1$), if some part of $S^-_L$ (denoted by ``$S^-_{Lp}$") exists in the set $\{(g_0,g_1): v_j\leq g_1 <v_{j+1},~g_0\geq s_{j}\}$ of ${\cal{R}}^*_{L1}$, we have $\frac{1}{2}\log(1+g_1p_j)\geq r_0$, $\frac{1}{2}\log(1+g_1p_{j+1})< r_0$ and $(\lambda+\mu g_0)(p_j-p_L)\geq 1$. And given $S^-_{Lp}\subset(\cup^{L-1}_{k=1} {\cal{R}}_k)$, let the power level for $S^-_{Lp}$ be $p_k$ (where $k$ could be any value from $\{1,\dots,L-1\}$). Reassigning this part of set
$S^-_L$ into region ${\cal{R}}_L$ will reduce the value of the LDF, since if $k\leq j$ (implying $p_k\geq p_j$), $g'(\lambda, \mu)-g(\lambda, \mu)=E[1+p_L(\lambda+\mu
g_0)-p_k(\lambda+\mu
g_0)|S^-_{Lp}]Pr (S^-_{Lp})<0$ and if $k>j$ (implying $p_k<p_j$), $g'(\lambda, \mu)-g(\lambda, \mu)=E[1+p_L(\lambda+\mu
g_0)-1-p_k(\lambda+\mu
g_0)|S^-_{Lp}]Pr (S^-_{Lp})<0$. 3) If some part of $S^-_L$ belongs
to the set ${\cal{R}}^*_{L2}$, similar to
(a), we can show that the new partition for this part of $S^-_L$
does not change the overall outage probability and meanwhile
reduces the value of the LDF. These all contradict  
optimality.\\
{\bf (2)}: We will now show that the set $S^+_j=\emptyset,~
j=1,\dots,L$. When $j=L$, it's straightforward that
$S^+_L=\emptyset$. When $j<L$, we assume that
$S^+_j\not=\emptyset$. Within the set $S^+_j\in {\cal{R}}^*_j$, we
have $v_j\leq g_1 <v_{j+1}$, implying
$\frac{1}{2}\log(1+g_1p_{j+1})< r_0$, or in other words,
$S^+_j\in(\cup^{L}_{k=j+1}{\cal{R}}_k)$ is in outage. We can
reallocate the set $S^+_j$ into region ${\cal{R}}_j$. This
reassignment not only lowers the outage probability ($S^+_j$ with
$p_j$ will not be in outage) but also lowers the value of the LDF,
given by $g'(\lambda, \mu)-g(\lambda, \mu)=E[(\lambda+\mu
g_0)(p_j-p_k)-1|S^+_j]Pr (S^+_j)\leq E[(\lambda+\mu
g_0)(p_j-p_L)-1|S^+_j]Pr (S^+_j)<0$, due to
$g_0<s_j=\frac{1}{\mu(p_j-p_L)}-\frac{\lambda}{\mu}$. This also
contradicts optimality. \\
\indent Therefore, we have ${\cal{R}}^*_j \setminus {\cal{R}}_j=\emptyset, ~\forall j=1,\dots,L$, i.e, ${\cal{R}}^*_j \subseteq {\cal{R}}_j, ~\forall j=1,\dots,L$. Since $\cup^{L}_{j=1}{\cal{R}}^*_j=\text{the whole space of}~(g_0, g_1)=\cup^{L}_{j=1}{\cal{R}}_j$, and ${\cal{R}}^*_j \subseteq {\cal{R}}_j,~\forall j$, we can obtain that ${\cal{R}}^*_j = {\cal{R}}_j,~\forall j=1,\dots,L.$
\subsection{Proof of Lemma \ref{ul2}:}\label{app4}
 We assume that $\lim_{L\rightarrow\infty}p_L\not=0$. Let $\delta=\lim_{L\rightarrow\infty}p_L> 0$. From the KKT condition (\ref{ukkt2}), we have
\begin{align}
e^{-v_L}\frac{c}{p^2_L}(p_{L-1}-p_L)(\lambda+\mu(1-e^{-s_{L-1}}))&= (\lambda+\mu)(P^L_{out}+e^{-v_L})+\mu \sum^{L-1}_{j=1}(e^{-v_j}-e^{-v_{j+1}})e^{-s_j}s_j\nonumber\\
&\geq (\lambda+\mu)(P^L_{out}+e^{-v_L})
\label{uk70}
\end{align}
Let $P^f_{out}$ denote the outage probability with full CSI at SU-TX, then we have $P^L_{out}\geq P^f_{out}$  and $\lim_{L\rightarrow\infty}P^L_{out}=P^f_{out}$. Taking the limit $L\rightarrow\infty$ on both sides of (\ref{uk70}), we have
\begin{align}
\lim_{L\rightarrow\infty} e^{-v_L}\frac{c}{p^2_L}(p_{L-1}-p_L)(\lambda+\mu(1-e^{-s_{L-1}}))\geq (\lambda+\mu)(P^f_{out}+e^{-\frac{c}{\delta}})\not=0
\label{uk71}
\end{align}
Given $p_1>\dots>p_L>0$, it is clear that the sequence $\{p_j\},\; j=1,2, \ldots, L$ is a monotonically decreasing sequence bounded below, therefore it must converge to its greatest-lower bound $\delta$, as $L \rightarrow \infty$.
%(any bounded monotonically decreasing sequence is convergent and the limit is its greatest-lower bound ).
Therefore, it can be easily shown that for an arbitrarily small
$\epsilon > 0$, we  can always find a sufficiently large $L$ such
that $p_{L-1} - p_L < \epsilon$. Thus, as $L\rightarrow \infty$,
$(p_{L-1}-p_L)\rightarrow 0$, which implies when $\mu>0$,
$s_{L-1}=\frac{1}{\mu(p_{L-1}-p_L)}-\frac{\lambda}{\mu}\rightarrow
\infty$. This implies that
\begin{align}
\lim_{L\rightarrow\infty}e^{-v_L}\frac{c}{p^2_L}(p_{L-1}-p_L)(\lambda+\mu(1-e^{-s_{L-1}}))= e^{-\frac{c}{\delta}}\frac{c}{\delta^2}(\lambda+\mu) \lim_{L\rightarrow\infty}(p_{L-1}-p_L)=0.
\end{align}
which is in contradiction with (\ref{uk71}). Thus, we must have
$\lim_{L\rightarrow\infty}p_L=0$.

\subsection{Proof of Lemma \ref{ul3}:}\label{app1}
 As $L\rightarrow\infty$, from Lemma \ref{ul2}, we have $p_L\rightarrow 0$. Applying it to Problem (\ref{uQ3}), we have the KKT conditions as (\ref{ukkt6}).\\
\indent 1) $\mu>0$:  From $s'_j=\frac{1}{\mu p_j}-\frac{\lambda}{\mu}$, we have
$p_j=\frac{1}{\lambda+\mu s'_j}$, and we also have
$p_0=\frac{1}{\lambda+\mu s'_0}$.  Applying it to (\ref {ukkt6}),
the right hand side (RHS) of equation (\ref {ukkt6}) becomes,
\begin{align}
RHS&=e^{-v_j}\frac{c}{p^2_j}[\frac{\lambda+\mu(1-e^{-s'_{j-1}})}{\lambda+\mu s'_{j-1}}-\frac{\lambda+\mu(1-e^{-s'_{j}})}{\lambda+\mu s'_j}]\nonumber\\
&=e^{-v_j}\frac{c(s'_{j-1}-s'_j)}{p^2_j}\frac{\frac{\lambda+\mu(1-e^{-s'_{j-1}})}{\lambda+\mu
s'_{j-1}}-\frac{\lambda+\mu(1-e^{-s'_{j}})}{\lambda+\mu
s'_j}}{s'_{j-1}-s'_j} \label{uk1}
\end{align}
From the mean value theorem (MVT), we have
\begin{align}
\frac{\frac{\lambda+\mu(1-e^{-s'_{j-1}})}{\lambda+\mu s'_{j-1}}-\frac{\lambda+\mu(1-e^{-s'_{j}})}{\lambda+\mu s'_j}}{s'_{j-1}-s'_j}
=\frac{-\mu}{(\lambda+\mu
s')^2}[\lambda(1-e^{-s'})+\mu(1-e^{-s'}(1+s'))] \label{uk2}
\end{align}
where $s'\in [s'_{j-1}, s'_{j})$. As the number of feedback bits
$B=\log_2 L\rightarrow \infty$, the length of quantization interval
on $g_0$ axis $[s'_{j-1}, s'_{j}), j=1,\dots,L-1$ approaches zero
\cite{icc}. Hence (\ref{uk2}) becomes,
\begin{align}
\frac{\frac{\lambda+\mu(1-e^{-s'_{j-1}})}{\lambda+\mu s'_{j-1}}-\frac{\lambda+\mu(1-e^{-s'_{j}})}{\lambda+\mu s'_j}}{s'_{j-1}-s'_j}
\approx \frac{-\mu}{(\lambda+\mu
s'_j)^2}[\lambda(1-e^{-s'_j})+\mu(1-e^{-s'_j}(1+s'_j))]
\label{uk3}
\end{align}
Applying (\ref{uk3}) to (\ref{uk1}), we have
$RHS \approx
e^{-v_j}c\mu(s'_{j}-s'_{j-1})[\lambda(1-e^{-s'_j})+\mu(1-e^{-s'_j}(1+s'_j))]$.
Similarly, as $L\rightarrow \infty$, we also have the length of
quantization interval on the $g_1$ axis $[v_{j}, v_{j+1}),
j=1,\dots,L-2$ approaches zero, thus from MVT,
$e^{-v_j}-e^{-v_{j+1}}\approx e^{-v_j}(v_{j+1}-v_j)$.
Thus the left hand side (LHS) of equation (\ref {ukkt6}) can be
approximated as,
$ LHS\approx
e^{-v_j}(v_{j+1}-v_j)[\lambda(1-e^{-s'_j})+\mu(1-e^{-s'_j}(1+s'_j))$.
Hence, we have $\forall j=1,\dots,L-2,$
$v_{j+1}-v_j\approx c\mu (s'_{j}-s'_{j-1})$, from which we get
$s'_{j+1}-s'_j\approx  s'_{j}-s'_{j-1}, \forall j=1,\dots,L-2$,
namely,
$s'_{L-1}-s'_{L-2}\approx\dots\approx  s'_{1}-s'_{0}$,
since $v_j=c\lambda+c\mu s'_{j}$.

2) $\mu=0$:  In this case, we have $s'_j=\infty, j=1,\dots,L-1$. Thus (\ref{ukkt6})
becomes
$e^{-v_j}-e^{-v_{j+1}}=e^{-v_j}\frac{c}{p^2_j}(p_{j-1}-p_j)$,
where $j=1,\dots,L-1$ and $p_0=\frac{1}{\lambda}$, which
can be rewritten as
$\frac{1}{v_j} (e^{-v_j}-e^{-v_{j+1}})=\frac{1}{v_{j-1}}
e^{-v_j}(v_{j}-v_{j-1})$,
where $v_0=\frac{c}{p_0}=c\lambda$. Applying MVT into
as before, we have
$\frac{1}{v_j} e^{-v_j}(v_{j+1}-v_j)\approx\frac{1}{v_{j-1}}
e^{-v_j}(v_{j}-v_{j-1}), \forall j=1,\dots,L-2$,
which yields
$\frac{v_{j+1}}{v_j}\approx\frac{v_{j}}{v_{j-1}}, \forall j=1,\dots,L-2$,
namely,
$\frac{v_{L-1}}{v_{L-2}}\approx\dots\approx\frac{v_{1}}{v_{0}}$.

This completes the proof for Lemma \ref{ul3}.
\subsection{Proof of Lemma \ref{ul4}:}\label{app2}
 As $L\rightarrow\infty$, from
Lemma \ref{ul2}, we have $p_L\rightarrow 0$. Adding the two
equations of (\ref{ukkt3}) together and applying $p_L\rightarrow
0$, we have
\begin{align}
\sum^{L-1}_{j=1} p_j (e^{-v_j}-e^{-v_{j+1}})[\lambda (1-e^{-s'_j})+\mu (1-e^{-s'_j}(1+s'_j))]=\lambda P_{av}+\mu Q_{av} \label{uk21}
\end{align}
The KKT conditions (\ref{ukkt6}) can be rewritten as
\begin{align}
p_j(e^{-v_j}-e^{-v_{j+1}})[\lambda(1-e^{-s'_j})+\mu(1-e^{-s'_j}(1+s'_j))]
=p_{j-1}e^{-v_j}(v_j-v_{j-1})\frac{[\hat{f}'(p_{j-1})-\hat{f}'(p_j)]}{p_{j-1}-p_j}
\label{uk22}
\end{align}
where $\hat{f}'(p_j)=p_j(\lambda+\mu(1-e^{-s'_{j}}))$. As
mentioned before, when $L\rightarrow \infty$, we have the length of
quantization interval on the $g_1$ axis $[v_{j-1}, v_{j}),
j=2,\dots,L-1$ approaching zero. Hence we also have the length of the interval $[p_{j-1},
p_{j}), j=2,\dots,L-1$ approaching zero, since
$v_j=\frac{c}{p_j}$. Thus from MVT, we have
\begin{align}
e^{-v_{j-1}}-e^{-v_j}&\approx e^{-v_j}(v_j-v_{j-1})\nonumber\\
\frac{\hat{f}'(p_{j-1})-\hat{f}'(p_j)}{p_{j-1}-p_j}&\approx\lambda(1-e^{-s'_{j-1}})+\mu(1-e^{-s'_{j-1}}(1+s'_{j-1}))
\label{uk23}
\end{align}
Applying (\ref{uk23}) into (\ref{uk22}), we can obtain,
$\forall j=2,\dots,L-1$
\begin{align}
&p_j(e^{-v_j}-e^{-v_{j+1}})[\lambda(1-e^{-s'_j})+\mu(1-e^{-s'_j}(1+s'_j))]\approx\nonumber\\
&p_{j-1}(e^{-v_{j-1}}-e^{-v_j})[\lambda(1-e^{-s'_{j-1}})+\mu(1-e^{-s'_{j-1}}(1+s'_{j-1}))]
\label{uk24}
\end{align}
Then applying the result of (\ref{uk24}) into (\ref{uk21}), we can
have $j=1,\dots,L-1$
\begin{align}
p_j(e^{-v_j}-e^{-v_{j+1}})[\lambda(1-e^{-s'_j})+\mu(1-e^{-s'_j}(1+s'_j))]\approx \frac{\lambda P_{av}+\mu Q_{av}}{L-1} \label{uk25}
\end{align}
which gives,
\begin{align}
\sum^{L-1}_{j=1} (e^{-v_j}-e^{-v_{j+1}})[\lambda(1-e^{-s'_j})+\mu(1-e^{-s'_j}(1+s'_j))]\approx \frac{\lambda P_{av}+\mu Q_{av}}{L-1} \sum^{L-1}_{j=1}
\frac{1}{p_j}. \label{uk26}
\end{align}
This completes the proof for Lemma \ref{ul4}.
\subsection{Proof of Theorem \ref{ut1}:}\label{app3}
 1)  $\mu>0$:  From  Lemma \ref{ul3}, we can easily
obtain,
$s'_{j} \approx js'_{1},\;
\frac{1}{p_j}=\lambda+\mu s'_{j}\approx \lambda+j\mu s'_{1}$, and
${v_j}=\frac{c}{p_j}\approx c\lambda+jc\mu s'_{1}, \forall j=1,\dots,L-1$,
Let
$z=\sum^{L-1}_{j=1} (e^{-v_j}-e^{-v_{j+1}})[\lambda(1-e^{-s'_j})+\mu(1-e^{-s'_j}(1+s'_j))]$,
which implies that $0<z<\lambda+\mu$. Then from Lemma \ref{ul4},
we have
$\frac{1}{L-1} \sum^{L-1}_{j=1} \frac{1}{p_j}\approx z'$,
where $z'=\frac{z}{\lambda P_{av}+\mu Q_{av}}$ and
$0<z'<\frac{\lambda+\mu}{\lambda P_{av}+\mu Q_{av}}$. Using the above results, we get
$s'_{1}\approx \frac{2(z'-\lambda)}{\mu L}=\frac{d}{L}$,
where $d=\frac{2(z'-\lambda)}{\mu}$. Let $a=c\mu
d=2(z'-\lambda)c$, then $s'_{1} \approx \frac{a}{c\mu L}$. Since
$0<z'<\frac{\lambda+\mu}{\lambda P_{av}+\mu Q_{av}}$, we have
$\lim_{L\rightarrow \infty}\frac{a}{L}=0$.
From the definition of $z$ above, we have
\begin{align}
z&=(\lambda+\mu)e^{-v_1}-\sum^{L-1}_{j=1} (e^{-v_j}-e^{-v_{j+1}})[(\lambda+\mu) e^{-s'_j}+\mu e^{-s'_j}s'_j]\nonumber\\
&\approx
e^{-c\lambda}[(\lambda+\mu)e^{-\frac{a}{L}}-(1-e^{-\frac{a}{L}})(\lambda+\mu)\sum^{L-1}_{j=1}
e^{-j(\frac{a}{L}+s'_1)}-(1-e^{-\frac{a}{L}})\mu s'_1 \sum^{L-1}_{j=1} je^{-j(\frac{a}{L}+s'_1)}]\nonumber\\
&\approx
e^{-c\lambda}[(\lambda+\mu)e^{-\frac{a}{L}}-(1-e^{-\frac{a}{L}})(\lambda+\mu)\sum^{L-1}_{j=1}
e^{-j\frac{b}{L}}-(1-e^{-\frac{a}{L}})\frac{a}{c L} \sum^{L-1}_{j=1}
je^{-j\frac{b}{L}}]
\label{uk33}
\end{align}
where $b=a+Ls'_1=a(1+\frac{1}{c\mu})$ and we also have
$\lim_{L\rightarrow \infty}\frac{b}{L}=0$. Since
$\sum^{L-1}_{j=1} e^{-j\frac{b}{L}} =\frac{1-e^{-b}}{1-e^{-\frac{b}{L}}}-1$, and
$\sum^{L-1}_{j=1}
je^{-j\frac{b}{L}} =-\frac{e^{(-\frac{b}{L}-b)}(Le^{\frac{b}{L}}-e^b-L+1)}{(1-e^{-\frac{b}{L}})^2}$,
(\ref{uk33}) becomes
\begin{align}
z & \approx
e^{-c\lambda}[(\lambda+\mu)(1-(1-e^{-\frac{a}{L}})\frac{1-e^{-b}}{1-e^{-\frac{b}{L}}})-(1-e^{-\frac{a}{L}})\frac{a}{c
L}\frac{e^{-\frac{b}{L}}(1-e^{-b})-Le^{-b}(1-e^{-\frac{b}{L}})}{(1-e^{-\frac{b}{L}})^2}
] \label{uk34}
\end{align}
Since $\lim_{L\rightarrow \infty}\frac{a}{L}=0$ and
$\lim_{L\rightarrow \infty}\frac{b}{L}=0$, we have
$1-e^{-\frac{a}{L}}\approx \frac{a}{L}$ and
$1-e^{-\frac{b}{L}}\approx \frac{b}{L}$.
And when $L\rightarrow \infty$, we approach the full CSI scenario, thus implying
$\lambda \approx \lambda^f,
\mu \approx \mu^f$.
Using these results in (\ref{uk34}),  we have
\begin{align}
z&\approx
e^{-c\lambda_f^*}[(\lambda_f^*+\mu_f^*)(1-\frac{a}{b}(1-e^{-b}))-\frac{a^2}{cb^2
}((1-\frac{b}{L})(1-e^{-b})-be^{-b}) ]\nonumber\\
&\approx
e^{-c\lambda_f^*}[(\lambda_f^*+\mu_f^*)(1-\frac{a}{b}(1-e^{-b}))-\frac{a^2}{cb^2
}(1-e^{-b}(1+b))]
\label{uk36}
\end{align}
%\nonumber\\
%&=e^{-c\lambda}[(\lambda+\mu)(1-\frac{c\mu}{1+c\mu}(1-e^{-a(1+\frac{1}{c\mu})}))-\frac{c
%\mu^2}{(1+c\mu)^2
%}(1-e^{-a(1+\frac{1}{c\mu})}(1+a(1+\frac{1}{c\mu})))]
Since
$z=(\lambda P_{av}+\mu Q_{av})z'=(\lambda P_{av}+\mu
Q_{av})(\lambda+\frac{a}{2c})\approx(\lambda_f^* P_{av}+\mu_f^*
Q_{av})(\lambda_f^*+\frac{a}{2c})$,  we can obtain $a$ from the following approximation:
\begin{align}
&(\lambda_f^* P_{av}+\mu_f^* Q_{av})(\lambda_f^*+\frac{a}{2c})e^{c\lambda_f^*}\nonumber\\
&\approx [(\lambda_f^*+\mu_f^*)(1-\frac{c\mu_f^*}{1+c\mu_f^*}(1-e^{-a(1+\frac{1}{c\mu_f^*})}))-\frac{c (\mu_f^*)^2}{(1+c\mu_f^*)^2
}(1-e^{-a(1+\frac{1}{c\mu_f^*})}(1+a(1+\frac{1}{c\mu_f^*})))]
\label{uk37}
\end{align}
From (\ref{uk37}), with given $P_{av}$ and $Q_{av}$, $a$ is
a constant. Then when $L$ is large,
\begin{align}
P^L_{out}&\approx
1-e^{-v_1}+\sum^{L-1}_{j=1}(e^{-v_j}-e^{-v_{j+1}})e^{-s'_j}
\approx
1-e^{-c\lambda}[e^{-\frac{a}{L}}-(1-e^{-\frac{a}{L}})\sum^{L-1}_{j=1}
e^{-j\frac{b}{L}}]\nonumber\\
 &=
1-e^{-c\lambda}[1-(1-e^{-\frac{a}{L}})\frac{1-e^{-b}}{1-e^{-\frac{b}{L}}}]
 \approx
1-e^{-c\lambda_f^*}[1-(1-e^{-\frac{a}{L}})\frac{1-e^{-a(1+\frac{1}{c\mu_f^*})}}{1-e^{-\frac{a(1+\frac{1}{c\mu_f^*})}{L}}}]
\end{align}
and
$\lim_{L\rightarrow \infty} P^L_{out}=
1-e^{-c\lambda_f^*}[1-\frac{1-e^{-a(1+\frac{1}{c\mu_f^*})}}{1+\frac{1}{c\mu_f^*}}]$. \\
\indent 2)  $\mu=0$:   Let $y=\frac{v_1}{v_0}=\frac{v_1}{c\lambda}>1$, then again, from Lemma \ref{ul3}, we can get for $j=1,\dots,L-1$,
$v_{j} \approx c\lambda y^j$.
From Lemma \ref{ul4}, we have
$e^{-c\lambda y}\approx \frac{\lambda
P_{av}}{L-1}\sum^{L-1}_{j=1}y^j=\frac{\lambda
P_{av}}{L-1}\frac{y^L-y}{y-1}$.
With $x=y-1$, we have,
$e^{-c\lambda (1+x)}\approx {\lambda
P_{av}}(1+x)\frac{(1+x)^{L-1}-1}{x(L-1)}$.
Now, suppose  $\lim_{L\rightarrow\infty} xL=\infty$. Since
$(1+x)^{L-1}> 1+(L-1)x+\frac{1}{2}(L-2)(L-1)x^2$,
we have
$\lim_{L\rightarrow\infty}\frac{(1+x)^{L-1}-1}{(L-1)x}>
\lim_{L\rightarrow\infty} 1+\frac{1}{2}(L-2)x=\infty$.
Then taking the limit as $L\rightarrow\infty$, we have
$\lim_{L\rightarrow\infty} e^{-c\lambda (1+x)}=\infty$,
which contradicts $\lim_{L\rightarrow\infty} e^{-c\lambda
(1+x)}< 1$, thus we must have $\lim_{L\rightarrow\infty}
xL= 0 \leq \beta<\infty$ (where $\beta$ is a constant), implying
as $L\rightarrow\infty$, $x \rightarrow
\frac{\beta}{L}$.
Applying this result, we get
$e^{-c\lambda (1+\frac{\beta}{L})}\approx {\lambda
P_{av}}(1+\frac{\beta}{L})\frac{(1+\frac{\beta}{L})^{L-1}-1}{\frac{\beta}{L}(L-1)}$.
After taking the limit as $L\rightarrow\infty$ on  both
sides of above equation, we have
$e^{-c\lambda_f^*}\approx {\lambda_f^* P_{av}}\frac{e^{\beta}-1}{\beta}$, from which one can solve
for $\beta$ approximately.
Note that in the above approximation, we have used
$\lim_{L\rightarrow\infty}(1+\frac{\beta}{L})^{L-1}=e^{\beta}$ and
 when $L$ is large, $\lambda\approx\lambda_f^*$.
Therefore, when $L$ is large,
$P^L_{out} =1-e^{-v_1}= 1-e^{-c\lambda (1+x)}\approx 1-e^{-c\lambda_f^* (1+\frac{\beta}{L})}$,
$\lim_{L\rightarrow \infty} P^L_{out}= 1-e^{-c\lambda_f^*}$.

This completes the proof for Theorem \ref{ut1}.
\end{appendix}
\renewcommand{\baselinestretch}{1}
\bibliographystyle{IEEEtran}

\newpage

\end{document}